\documentclass[prd,showpacs,twocolumn,
amsmath,amssymb,superscriptaddress,floatfix,nofootinbib,10pt]{revtex4-1}
\usepackage[utf8]{inputenc}
\usepackage[sort&compress]{natbib}
\usepackage[normalem]{ulem}
\usepackage{lipsum} 
\usepackage{bm}
\usepackage{times}
\usepackage{amssymb,amsbsy,amsmath,amsfonts}
\usepackage{graphicx}
\usepackage{float}
\usepackage{color}
\usepackage{morefloats}
\usepackage{rotating}
\usepackage{srcltx}
\usepackage{slashed}
\usepackage{subfigure}
\usepackage{multirow}
\usepackage{verbatim}
\usepackage{hyperref}
\usepackage{tabularx}
\usepackage{adjustbox}
\usepackage[dvipsnames]{xcolor}
\usepackage{nicefrac}
\usepackage{soul}

\usepackage{hyperref}
\hypersetup{
	colorlinks	=true,
	urlcolor	=blue,
	linkcolor	=blue,
	citecolor	=blue,
	pdftitle	={},
	pdfauthor	={Zhi-Wei Liu, Jun-Xu Lu, Li-Sheng Geng},
	pdfsubject	={$DK$ correlation function}
	}

\begin{document}

\title{
Study of the $DK$ interaction with femtoscopic correlation functions
}

\author{Zhi-Wei Liu}
\affiliation{School of Physics, Beihang University, Beijing 102206, China}

\author{Jun-Xu Lu}
\affiliation{School of Space and Environment, Beijing 102206, China}
\affiliation{School of Physics, Beihang University, Beijing 102206, China}

\author{Li-Sheng Geng}
\email[Corresponding author: ]{lisheng.geng@buaa.edu.cn}
\affiliation{School of Physics,  Beihang University, Beijing 102206, China}
\affiliation{Peng Huanwu Collaborative Center for Research and Education, Beihang University, Beijing 100191, China}
\affiliation{Beijing Key Laboratory of Advanced Nuclear Materials and Physics, Beihang University, Beijing 102206, China }
\affiliation{Southern Center for Nuclear-Science Theory (SCNT), Institute of Modern Physics, Chinese Academy of Sciences, Huizhou 516000, Guangdong Province, China}

\begin{abstract}
The $DK$ interaction in isospin zero is known to be attractive to such an extent that a bound state can be generated, which can be associated with the mysterious $D_{s0}^*(2317)$. In this work, we calculate the $DK$  femtoscopic correlation function in the coupled-channel framework for different source sizes that  can directly probe the strongly attractive $DK$ interaction, which is otherwise inaccessible due to the unstable nature of $D$ and $K$ mesons,  and therefore can help elucidate the nature of $D_{s0}^*(2317)$.  We further generalize the study of source size dependence to various  interactions, ranging from repulsive, weakly attractive, moderately attractive, and strongly attractive, in a square-well model. We hope that our study can motivate  future experimental measurements of the $DK$ correlation function and other interactions relevant to the understanding of the nature of the many exotic hadrons discovered so far.  
\end{abstract}


\maketitle

\section{Introduction}

Since 2003, many of the so-called exotic hadrons that do not easily fit into the conventional quark model picture of $q\bar{q}$ mesons and $qqq$ baryons have been observed experimentally~\cite{Oset:2016lyh,Richard:2016eis,Hosaka:2016ypm,Chen:2016qju,Esposito:2016noz,Lebed:2016hpi,Guo:2017jvc,Ali:2017jda,Olsen:2017bmm,Karliner:2017qhf,Liu:2019zoy,vanBeveren:2020eis,Chen:2022asf,Mai:2022eur}, which opened a new era in hadron spectroscopy. Due to the fact that 
most (if not all) of such hadrons are located close to the thresholds of two conventional hadrons, they are often interpreted as hadronic molecules~\cite{Guo:2017jvc}, similar to the deuteron which can be viewed as a bound state of a proton and a neutron~\cite{Weinberg:1965zz}. Nevertheless, how to verify the molecular picture remains a challenging task both experimentally and theoretically.

Many attempts have been made to address this challenge from various perspectives. For instance, inspired by the formation of atomic nuclei from clusters of nucleons bound by the nuclear force, the hadronic molecular picture can be checked by studying whether multi-body molecules can be formed by adding a third hadron to the two-body bound states~\cite{Wu:2022ftm,Wu:2021dwy,MartinezTorres:2020hus}. Alternatively, the production yields of exotic multiquark states in high-energy collisions, which are expected to be strongly affected by their internal structure, have received increasing attention~\cite{Cho2011PRL106.212001,ExHIC:2017smd, Zhang2021PRL126.012301, Hu2021PRD104.L111502, Chen2022PRD105.054013,Esposito:2020ywk,Chen:2021akx,CMS:2021znk,Xu:2021drf}. In a recent work~\cite{Wu2022arXiv2211.01846}, it was shown for the first time that the production yields of $D_{s0}^*(2317)$ and $D_{s1}(2460)$ measured by the BaBar experiment~\cite{BaBar:2006eep} can   be reproduced in the molecular picture in a model independent way. Nevertheless, direct evidence for the attractive strong interactions  responsible for the formation of hadronic molecules is still lacking.

In recent years, Femtoscopy, which measures two-particle momentum correlation functions in high-energy proton-proton ($pp$), proton-nucleus ($pA$), and nucleus-nucleus ($AA$) collisions, has made remarkable progress in probing  the strong interactions between various pairs of hadrons~\cite{STAR2015Nature527.345, STAR2015PRL114.022301, HADES2016PRC94.025201, ALICE2017PLB774.64, ALICE2017PRC96.064613, ALICE2019PRL123.112002, ALICE2019PLB797.134822, ALICE2019PLB790.22, STAR2019PLB790.490, ALICE2019PRC99.024001, ALICE2020Nature588.232, ALICE2020PRL124.092301, ALICE2020PLB802.135223, ALICE2020PLB805.135419, ALICE2021PRL127.172301, ALICE2021PLB813.136030, ALICE2021PRC103.055201, ALICE2022PLB829.137060, ALICE2022PLB833.137272, ALICE2022PLB833.137335, ALICE2022PRD106.052010}. In particular,  it was shown that correlation functions can help reveal the existence  of bound states~\cite{Ohnishi2017PPNP95.279, ALICE2021ARNPS71.377}.  In this sense, Femtoscopy offers a valuable means to directly test the hadronic molecular picture. It is worthwhile to note that recently Kamiya et al. studied the $DD^*$ and $D\bar{D}^*$~\footnote{Charge conjugated states are always implied unless otherwise stated.} correlation functions. They showed how the nature of $T_{cc}^+(3875)$ and $X(3872)$ as bound states of $DD^*$ and $D\bar{D}^*$  can be verified~\cite{Kamiya2022EPJA58.131}.

In the present work, we  study whether Femtoscopy  can help reveal the internal structure of more deep bound hadronic molecules. In the heavy quark sector, one of the most studied of such exotic hadrons is the $D_{s0}^*(2317)$ state (and its heavy-quark spin symmetry partner $D_{s1}(2460)$), which was discovered by the BaBar Collaboration in the inclusive $D_s^+\pi^0$ invariant mass distribution and subsequently confirmed by the CLEO and Belle Collaborations~\cite{BABAR2003PRL90.242001, CLEO2003PRD68.032002, Belle2003PRL91.262002}. Due to the fact the $D_{s0}^*(2317)$ is located below the $DK$ threshold by about $45$ MeV (lighter than the Godfrey-Isgur model by 160 MeV~\cite{Godfrey:1985xj, Godfrey:1986wj}), the prevailing picture is that it is largely a bound state of $DK$ and coupled channels~\cite{Beveren2003PRL91.012003, Barnes2003PRD68.054006, Chen2004PRL93.232001, Kolomeitsev2004PLB582.39, Gamermann:2006nm, Guo2007PLB647.133, Yang2022PRL128.112001,Liu:2022dmm}, which is supported by many lattice QCD simulations~\cite{Mohler2013PRL111.222001,Lang:2014yfa,Bali:2017pdv,Cheung:2020mql}.~\footnote{The only work claimed the contrary that we are aware of is Ref.~\cite{Alexandrou:2019tmk}.} Therefore, we take the $D_{s0}^*(2317)$ as a typical deep bound  state in the present work. In addition, we note that the ALICE Collaboration recently demonstrated that  Femtoscopy can be applied in the charm sector ~\cite{ALICE2022PRD106.052010, ALICE2022QM}.

The article is organized as follows. In Sec.~II we explain how to evaluate the $DK$ interaction and its correlation function. We  present the numerical results for the $DK$ correlation function and discuss the coupled-channel effects and the source size dependence in detail. In Sec. III, we generalize our study to the source size dependence of correlation functions for repulsive, weakly attractive, moderately attractive, and strongly attractive interactions, in the square-well model, which are relevant to  future studies of  interactions responsible for the formation of the many exotic hadrons discovered so far. We end with a short summary and outlook in Sec.IV.

\section{Femtoscopic study of the $DK$ interaction and the related $D_{s0}^*(2317)$}\label{subsection:DK}






In this section, we explain how to describe the $DK$ interaction and calculate the corresponding femtoscopic correlation function. To derive the interactions between a heavy pseudoscalar boson and a Nambu-Goldstone boson (NGB), we employ the leading order (LO) chiral perturbation theory (ChPT). In fact, this approach has successfully described the $DK$ interaction and the $D_{s0}^*(2317)$ at LO~\cite{Kolomeitsev:2003ac, Gamermann:2006nm, Guo2006PLB641.278}, NLO~\cite{Altenbuchinger2014PRD89.014026, Liu:2012zya, Guo:2015dha} and even NNLO~\cite{Yao:2015qia,Du:2017ttu,Huang:2022cag}. We note that the lattice QCD scattering lengths with unphysical light quark masses for the $DK$ and coupled channels~\cite{Liu:2012zya, Mohler:2013rwa} can also be well described in the NLO chiral perturbation theory~\cite{Altenbuchinger2014PRD89.014026, Liu:2012zya, Guo:2015dha}.
As a result, the chiral potential for the $DK$ interaction can be considered as well established. It should be noticed that in order to compare with future femtoscopy experiments, we work in the charge basis. The LO potential has the following Weinberg-Tomozawa form~\cite{Altenbuchinger2014PRD89.014026}:
\begin{align}
  V_{\nu'\nu}[D(p_1)\phi(p_2)&\rightarrow D(p_3)\phi(p_4)]\nonumber\\
  &=\frac{C_{\nu'\nu}}{4f_0^2}\left[(p_1+p_2)^2-(p_1-p_4)^2\right],
\end{align}
where $p_{1(3)}=(E_{1(3)},\boldsymbol{p^{(\prime)}})$ and $p_{2(4)}=(\sqrt{s}-E_{1(3)},-\boldsymbol{p^{(\prime)}})$ are the four-momenta of the incoming (outgoing) $D$ mesons and NGBs $\phi$. The initial (final) state index $\nu$ ($\nu'$) can be 1, 2, 3, and 4 which represent the  $D_s^+\pi^0$, $D^0K^+$, $D^+K^0$ and $D_s^+\eta$ coupled channels, respectively. The coefficients $C_{\nu'\nu}$ are listed in Table~\ref{Tab:Coefficients}. In addition, the meson masses $m_{D(\phi)}$ are taken from the latest review of particle physics (RPP)~\cite{PDG2022}, and the pion decay constant $f_0=92.2$ MeV. Because in the low-momentum region, the $S$-wave interaction is the most important, we only project the above potential to  $S$-wave:
\begin{align}
  V_{\nu'\nu}^{l=0}(\boldsymbol{p}',\boldsymbol{p};\sqrt{s})&=\frac{1}{2}\int_{-1}^1{\rm d}\cos\theta\nonumber\\
  &\times V_{\nu'\nu}\left[D(p_1)\phi(p_2)\rightarrow D(p_3)\phi(p_4)\right],
\end{align}
where $\theta$ represents the angle between the three-momenta of the initial and final states. Note that there is no free parameter in the LO potential.
\begin{table}[h]
  \centering
  \caption{Coefficients of the LO potential for $D\phi\rightarrow D\phi$.}
  \label{Tab:Coefficients}
  \setlength{\tabcolsep}{3pt}
  \begin{tabular}{cccccccccc}
    \hline
    \hline
    $C_{11}$&$C_{12}$ &$C_{13}$ &$C_{14}$ &$C_{22}$ &$C_{23}$ &$C_{24}$ &$C_{33}$ &$C_{34}$ &$C_{44}$ \\
    \hline
    $0$ &$1/\sqrt{2}$ &$-1/\sqrt{2}$ &$0$ &$-1$ &$-1$ &$\sqrt{3/2}$ &$-1$ &$\sqrt{3/2}$ &$0$ \\
    \hline
    \hline
  \end{tabular}
\end{table}

To calculate the femtoscopic correlation function, one needs the relative wave function of  the hadron pair of interests, determined by the two-particle interaction ~\cite{Ohnishi2017PPNP95.279, ALICE2021ARNPS71.377}. In general, the scattering wave function can be obtained by solving the Schr\"odinger equation in coordinate space~\cite{ALICE2018EPJC78.394, Ohnishi2021PRC105.014915}, or the Lippmann-Schwinger (Bethe-Salpeter) scattering equation in momentum space~\cite{Haidenbauer2019NPA981.1, Liu2022CPC}. As the ChPT potential is momentum-dependent (non-local),  for our purpose it is convenient to first obtain the reaction amplitude $T$ by solving the scattering equation $T=V+VGT$, and then derive the scattering wave function using the relation $|\psi\rangle=|\varphi\rangle+GT|\varphi\rangle$, where $G$ and $|\varphi\rangle$ represent the free propagator and the free wave function, respectively. More specifically, to obtain the reaction amplitude, we solve the following coupled-channel scattering equation (which is similar to the baryon-baryon case in Ref.~\cite{Liu2022CPC}),
\begin{align}\label{Eq:Kadyshevsky}
  T_{\nu'\nu}^{l=0}(\boldsymbol{p}',&\boldsymbol{p};\sqrt{s})=V_{\nu'\nu}^{l=0}(\boldsymbol{p}',\boldsymbol{p};\sqrt{s})+\sum_{\nu^{\prime\prime}}\int_0^\infty\frac{{\rm d}p^{\prime\prime}p^{\prime\prime2}}{8\pi^2}\nonumber\\
  &\times\frac{V_{\nu'\nu^{\prime\prime}}^{l=0}(\boldsymbol{p}',\boldsymbol{p}^{\prime\prime};\sqrt{s})\cdot T_{\nu^{\prime\prime}\nu}^{l=0}(\boldsymbol{p}^{\prime\prime},\boldsymbol{p};\sqrt{s})}{E_{D,\nu^{\prime\prime}}E_{\phi,\nu^{\prime\prime}}(\sqrt{s}-E_{D,\nu^{\prime\prime}}-E_{\phi,\nu^{\prime\prime}}+i\epsilon)},
\end{align}
where $E_{D(\phi),\nu^{\prime\prime}}=\sqrt{\boldsymbol{p}^{\prime\prime2}+m_{D(\phi),\nu^{\prime\prime}}^2}$. Here, in order to avoid ultraviolet divergence in numerical evaluations, we multiply the potential $V_{\nu'\nu^{(\prime\prime)}}^{l=0}$ of Eq.~\eqref{Eq:Kadyshevsky} by a simple Gaussian regulator to suppress high momentum contributions~\cite{Liu2019PRL122.242001},
\begin{align}\label{Eq:regularization}
  f_{\Lambda_F}(p,p')=\exp\left[-\left(\frac{p}{\Lambda_F}\right)^{2}-\left(\frac{p'}{\Lambda_F}\right)^{2}\right],
\end{align}
where $\Lambda_F$ is a cutoff parameter to be determined. Then with the half-off-shell $T$-matrix, we compute the $S$-wave scattering wave function in the following way
\begin{align}\label{Eq:Fourier_Bessel}
  \widetilde{\psi}_{\nu'\nu}^{l=0}(p,r)&=\delta_{\nu'\nu}j_{l=0}(pr)+\int_0^\infty\frac{{\rm d}p'p^{\prime2}}{8\pi^2}\nonumber\\
  &\times\frac{T_{\nu'\nu}^{l=0}(\boldsymbol{p}',\boldsymbol{p};\sqrt{s})\cdot j_{l=0}(p'r)}{E_{D,\nu^{\prime}}E_{\phi,\nu^{\prime}}(\sqrt{s}-E_{D,\nu^{\prime}}-E_{\phi,\nu^{\prime}}+i\epsilon)},
\end{align}
where $j_{l=0}$ is the spherical Bessel function for $l=0$. The above wave function is matched asymptotically to the boundary condition
\begin{align}\label{Eq:asymptote2}
\widetilde{\psi}_{\nu'\nu}^{l=0}(p_\nu,r)\overset{r\to\infty}{\longmapsto}&\frac{1}{2}\sqrt{\frac{\rho_{\nu'}}{\rho_\nu}}\bigg[\delta_{\nu'\nu}h_{l=0}^{(2)}(p_\nu r)+h_{l=0}^{(1)}(p_{\nu^\prime}r)\nonumber\\
&\times\left(\delta_{\nu'\nu}-2i\sqrt{\rho_{\nu'}\rho_\nu}\cdot T_{\nu'\nu}^{l=0}\right)\bigg],
\end{align}
where $h_{l=0}^{(1)}$ ($h_{l=0}^{(2)}$) represents the Hankel function of the first (second) kind for $l=0$, $T_{\nu'\nu}^{l=0}$ is the on-shell $T$-matrix, and the phase-space factor of channel $\nu$ is defined as $\rho_\nu=p_\nu/\left(8\pi\sqrt{s}\right)$ with the on-shell momentum $p_\nu$. As pointed out in Ref.~\cite{Haidenbauer2019NPA981.1}, one can recover the normalization of $\psi_{\nu'\nu}^{l=0}$ used for the correlation functions in Ref.~\cite{Ohnishi:2016elb} by multiplying $\widetilde{\psi}_{\nu'\nu}^{l=0}$ with the $S$-matrix $S_{\nu'\nu}=\delta_{\nu'\nu}-2i\sqrt{\rho_{\nu'}\rho_\nu}T_{\nu'\nu}$. This difference between $\psi_{\nu'\nu}^{l=0}$ and $\widetilde{\psi}_{\nu'\nu}^{l=0}$ is irrelevant as the $S$-matrix is unitary and one only needs the modulus squared of the wave function. To calculate the femtoscopic correlation function, one also needs the particle-emitting source created in relativistic heavy ion collisions~\cite{Ohnishi2017PPNP95.279, ALICE2021ARNPS71.377}. In this work, we adopt a common static and spherical Gaussian source with a single parameter $R$, namely, $S_{12}(r)=\exp[-r^2/(4R^2)]/(2\sqrt{\pi}R)^3$. With the aforementioned two theoretical ingredients, the correlation function can be calculated with the Koonin-Pratt (KP) formula~\cite{Koonin1977PLB70.43, Pratt1990PRC42.2646, Bauer1992ARNPS42.77}
\begin{align}\label{Eq:CF}
  C(p)=1+&\int_0^\infty4\pi r^2{\rm d}rS_{12}(r)\nonumber\\
  &\times\left[\sum_{\nu'}\omega_{\nu'}\left|\psi_{\nu'\nu}^{l=0}(p,r)\right|^2-\left|j_{l=0}(pr)\right|^2\right],
\end{align}
where $\omega_{\nu'}$ is the weight for each individual component of the multi-channel wave function, and the sum runs over all possible coupled channels. For simplicity we assume that the weights are the same and equal to $1$ in this exploratory study. For more details about the extension of the KP formula to the coupled-channel systems, we refer the reader to Refs.~\cite{Haidenbauer2019NPA981.1, Kamiya:2019uiw}.



As mentioned above, the cutoff $\Lambda_F$ is the only semi-free parameter in the description of the $DK$ scattering in the LO ChPT. It is fine-tuned to be $1107$ MeV so that the pole of the $T$-matrix is located at $\sqrt{s}=2317.8$ MeV~\cite{PDG2022} on the real axis, corresponding to the $D_{s0}^*(2317)$. The mass of $D_{s0}^*(2317)$, the threshold of the $D^0K^+$ pair, and the corresponding binding energy are listed in Table~\ref{Tab:ERE1}. In addition, for the sake of reference, we also calculate the scattering length and effective range that characterize the $D^0K^+$ interaction~\footnote{We adopt the ``nuclear physics'' convention for the scattering length, namely, $q\cot\delta=-1/a+r_{\rm eff}q^2/2+\mathcal{O}(q^4)$.}. Note that as discussed in Sec.~\ref{subsection:NM}, the effective range expansion up to $q^2$ does not work for the case of deep bound states studied here.
\begin{table}[h]
  \centering
  \caption{Binding energy $B$ (in units of MeV), scattering length $a$ and effective range $r_{\rm eff}$ (in units of fm) for the LO ChPT with $\Lambda_F=1107$ MeV. $D_{s0}^*(2317)$ mass and $D^0K^+$ threshold (in units of MeV) are also listed.}
  \label{Tab:ERE1}
  \setlength{\tabcolsep}{12.2pt}
  \begin{tabular}{ccccc}
    \hline
    \hline
    Mass  & Threshold  & $B~~$  & $a~~$  & $r_{\rm eff}$  \\ \hline
    $2317.8$  & $2358.52$  & $40.72$  & $0.73$  & $-2.11$  \\
    \hline
    \hline
  \end{tabular}
\end{table}

\begin{figure}[h]
  \centering
  \includegraphics[width=0.48\textwidth]{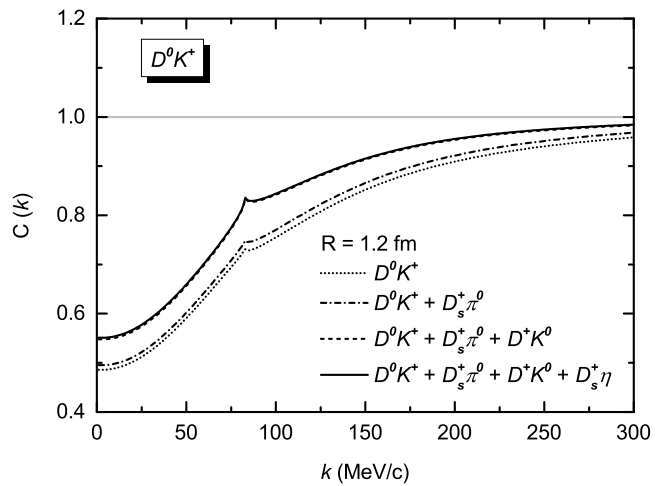}
  \caption{(color online) $D^0K^+$ correlation function as a function of the relative momentum $k$. The results are calculated with the LO chiral potential and a Gaussian source of $R=1.2$ fm. The dotted line denotes the correlation function for which only the $D^0K^+$ contribution is taken into account, while the dash-dotted line, the dashed line, and the solid line denote the results in which $(D_s^+\pi^0, D^+K^0, D_s^+\eta)$ contributions are considered one by one, respectively.
  }\label{Fig:D0Kp_CC}
\end{figure}

We present the $D^0K^+$ correlation function in Fig.~\ref{Fig:D0Kp_CC}. The results are obtained with the LO chiral potential and the Gaussian source ($R=1.2$ fm). It is seen that the $D^0K^+$ correlation is suppressed (compared to unity)  in a wide range of the relative momentum $k$, which features  the existence of a  bound state created from a large-size source or a repulsive interaction (see the discussions in Sec.III). Moreover, it is found that the inelastic coupled-channel effect is significant, which originates mainly from the $D^0K^+-D^+K^0$ transition. In particular, a cusp structure is seen at the $D^+K^0$ threshold ($k\simeq83$ MeV/c). Similar results are obtained for the $D^+K^0$ correlation function except that the cusp structure vanishes. Since the $D^+K^0$ threshold is slightly away from the $D_{s0}^*(2317)$ pole, the  strength of the $D^+K^0$ correlation function is slightly weaker than that of the $D^0K^+$ one.

\begin{figure}[h]
  \centering
  \includegraphics[width=0.48\textwidth]{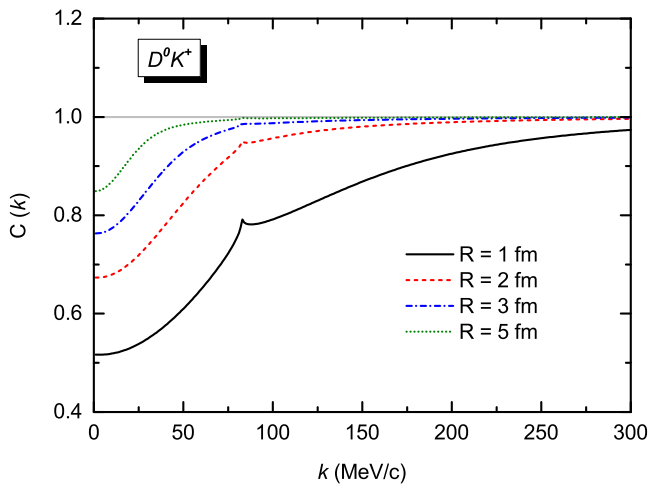}
  \caption{(color online) Source size dependence of the $D^0K^+$ correlation function. The results obtained with source sizes $R=1,2,3$, and $5$ fm are denoted by the black solid line, the red dashed line, the blue dash-dotted line, and the green dotted line, respectively.
  }\label{Fig:D0Kp_SS}
\end{figure}

The suppression of the correlation function for fixed $R$ alone cannot tell whether a shallow or deep bound state is present (given an attractive strong interaction). To draw a firm conclusion, we need to check the source size dependence of the $D^0K^+$ correlation function. As shown in Fig.~\ref{Fig:D0Kp_SS}, for both small and large collision systems, the $D^0K^+$ correlation functions are all between zero and unity, which is very different from the case of a moderately attractive interaction which only generates a shallow bound state (see discussions in Sec.III). In addition, the correlation strength decreases gradually with the increasing source size. This is understandable because of the short-range nature of the strong interaction. Note that the above results are closely related to the strongly attractive  $DK$ interaction, which is responsible for the bound state nature of the $D_{s0}^*(2317)$ state. Hence, to probe the $DK$ interaction in a model-independent way, we suggest that future experiments scan the $D^0K^+$ ($D^+K^0$) correlation functions by changing the source size in $pp$, $pA$, and $AA$ collisions, especially in the low-momentum region.

We note in passing that assuming the heavy-quark spin symmetry, we have also studied the $D^{*}K$ chiral potential and the corresponding correlation function relevant to $D_{s1}(2460)$. It is found that the $D^{*0}K^+$ correlation function and its source size dependence, as well as the coupled-channel effects, are  similar to those of the $D^0K^+$. The reason can be traced back to the fact that except for the difference between $D$ and $D^*$ meson masses, the strength and the structure of $DK$ and $D^{*}K$ interactions are almost the same in the leading order ChPT. One can expect similar results for the $B\bar{K}$ and $B^*\bar{K}$ correlation functions because of  the heavy-quark flavor symmetry~\cite{Guo2006PLB641.278, Altenbuchinger2014PRD89.014026}.

\section{Femtoscopic study of various  square-well potentials and their source size dependencies}\label{subsection:NM}

Having studied the $DK$ interaction,  we would like to discuss the above finding in a more general setting, and highlight some general features of correlation functions in relation to studies of hadronic molecules. For the sake of transparency and without loss of generality, we work with the square-well model and study four different  potentials, being either repulsive, weakly attractive, moderately attractive, or strongly attractive. The scattering wave function is obtained analytically by solving the stationary Schr\"odinger equation $-\frac{\hbar^2}{2\mu}\nabla^2\psi+V\psi=E\psi$ (considering only  $S$-wave interactions), where the reduced mass $\mu$ is chosen as $470$ MeV (close to that of the deuteron system). There are two parameters in the square-well potential $V(r)=V_0\theta(d-r)$, namely, the range parameter $d$ and the depth parameter $V_0$. In this work, $d$ is set at  $2.5$ fm, and $V_0$ is set at $25$, $-10$, $-25$, and $-75$ MeV for a repulsive potential, a weakly attractive potential not strong enough to generate a bound state, a moderately attractive potential capable of generating a shallow bound state, and a strongly attractive potential yielding a deep bound state, respectively. In the present work, we refer to a bound state as a shallow bound state if its binding energy  can be described by the effective-range expansion up to $q^2$, and  otherwise as a deep bound state. With the above parameters, the scattering lengths, the effective ranges, the approximate binding energies based on the effective-range expansion $B_{\rm ERE}=1/(2\mu r_{\rm eff}^2)\cdot\left(1-\sqrt{1-2r_{\rm eff}/a}\right)^2$~\cite{Gongyo:2017fjb, Naidon:2016dpf}, and the exact binding energies $B_{\rm Exact}$ are calculated and summarized in Table~\ref{Tab:ERE2}. As expected, a negative (positive) scattering length corresponds to the weakly attractive potential (repulsive potential or attractive potential capable of generating a bound state). In the case of the attractive potentials yielding a bound state, the scattering length decreases with the increasingly attractive interaction. It should be emphasized that in general the effective range is not equal to the range parameter of the square-well potential. In addition, the exact binding energy of the shallow bound state is $2.07$ MeV, which is close to the case of the deuteron, while the exact binding energy of the deep bound state is $38.08$ MeV, similar to that of $D_{s0}^*(2317)$ as a $DK$ bound state. We note that the approximate binding energy obtained from the effective-range expansion formula agrees well with the exact binding energy for the shallow bound state, but deviates significantly from the exact binding energy for the deep bound state, which is consistent with the definition of  shallow and deep bound states adopted in this work.

\begin{table}[h]
  \centering
  \caption{Scattering lengths $a$, effective ranges $r_{\rm eff}$ (in units of fm), approximate binding energies based on the effective-range expansion  $B_{\rm ERE}$ and exact binding energies $B_{\rm Exact}$ (in units of MeV) for the different square-well potentials.}
  \label{Tab:ERE2}
  \setlength{\tabcolsep}{4pt}
  \begin{tabular}{crrccc}
    \hline
    \hline
    Model  & $a~~$  & $r_{\rm eff}$  & $B_{\rm ERE}~~$  & $B_{\rm Exact}~~$  \\ \hline
    repulsive  & $1.27$  & $0.56$  & $-~~$  & $-~~$  \\
    weakly attractive  & $-3.24$  & $3.28$  & $-~~$  & $-~~$  \\
    moderately attractive  & $5.77$  & $2.06$  & $2.10$  & $2.07$  \\
    strongly attractive  & $2.33$  & $1.30$  & $21.44+16.79{\rm i}$  & $38.08$  \\
    \hline
    \hline
  \end{tabular}
\end{table}

\begin{figure*}[h]
  \centering
  \includegraphics[width=1\textwidth]{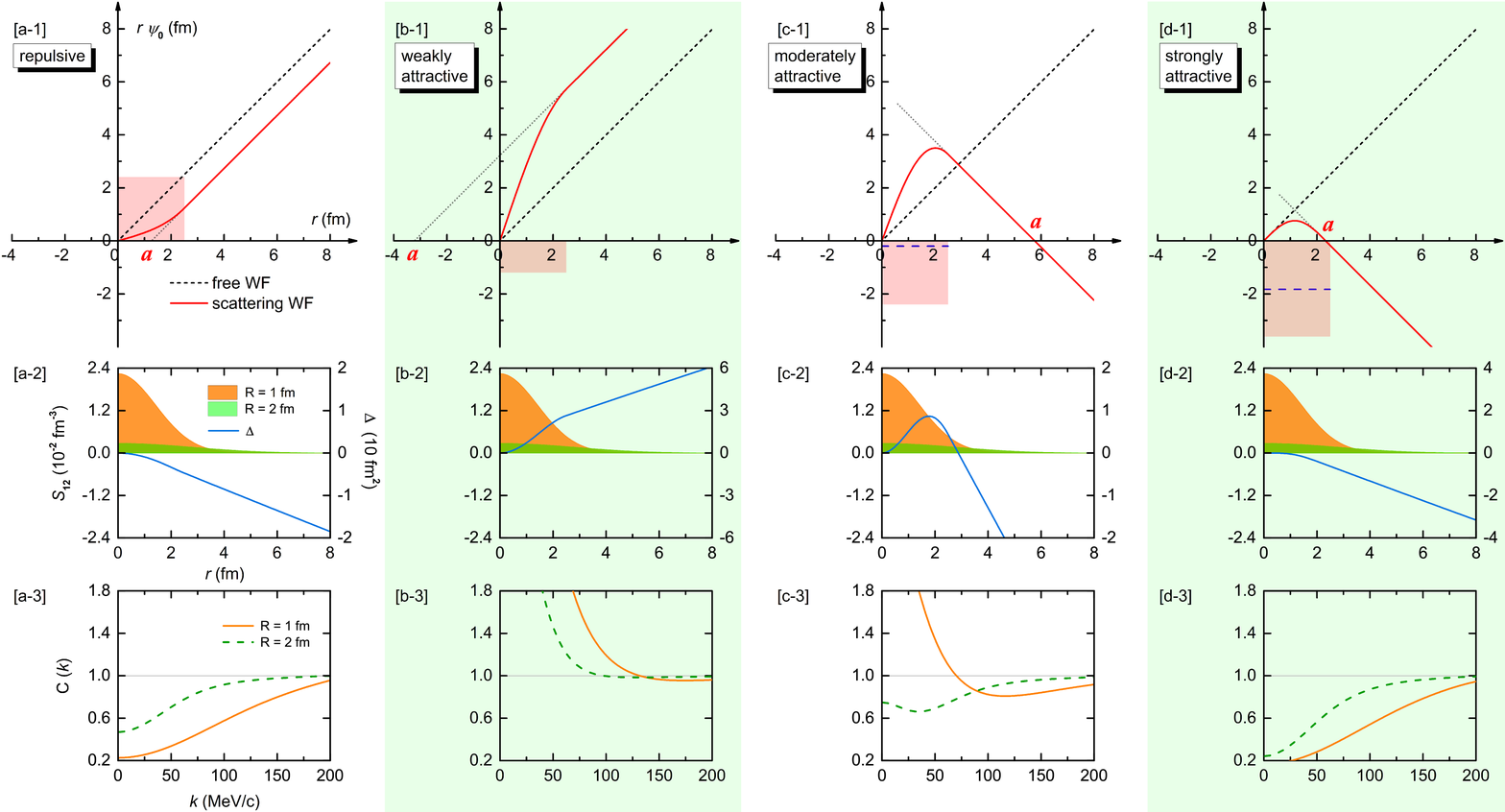}
  \caption{(color online) Scattering wave functions, source functions, and correlation functions for the four different square-well potentials, namely, (a) a repulsive potential, (b) a weakly attractive potential, (c) a moderately attractive potential, and (d) a strongly attractive potential.Panels (a1--d1): the product of the relative distance $r$ and the $S$-channel wave function $\psi_0$ as a function of $r$ in the low-momentum region. The black short-dashed line denotes the free-wave result, while the red solid line denotes the scattering-wave result. The light red region depicts the square-well potential. Panels (a2--d2): Gaussian source function $S_{12}$ as a function of $r$. The results are calculated with  a source size $R=1$ fm (dark orange region) and $R=2$ fm (light green region). In the same plot, the difference between the  free and scattering wave functions squared $\Delta\equiv r^2(|\psi_{l=0}|^2-|j_{l=0}|^2)$ in Eq.~\eqref{Eq:CF} is also shown as the blue solid line. Panels (a3--d3): Correlation function as a function of the relative momentum $k$. The orange solid line and the green dashed line denote the results obtained with $R=1$ fm and $R=2$ fm, respectively.  
  }\label{Fig:NM}
\end{figure*}
The general features of the correlation functions are shown in Fig.~\ref{Fig:NM}. Panels (a1-d1) of Fig.~\ref{Fig:NM} show the products of the relative distance $r$ and the $S$-channel wave function $\psi_0$ for the relative momentum $k\simeq3$ MeV/c which corresponds to $E=0.01$ MeV~\footnote{Note that this momentum is chosen as a representative case for illustration purposes only but without the loss of generality. The correlation functions shown in panels (a3-d3) are obtained with the scattering wave functions calculated at the corresponding relative momentum.}.  The results obtained with free and scattering wave functions are denoted by the red solid and black short-dashed lines, respectively. Here, the free-wave  result is equal to $r\cdot j_{l=0}$, which is a straight line with a slope of $1$ in the low-energy limit. Compared to the free-wave case, one can see that (a) for the repulsive potential the product is suppressed, (b) for the  weakly attractive potential the product is enhanced, (c) for the moderately attractive potential the product is enhanced at the short distance while suppressed at the long distance, and (d) for the strongly attractive potential the product is suppressed. It is interesting to note that panels (a1-d1) give the geometrical interpretations of the scattering lengths $a$ for the four potentials. In the square-well model, the scattering length $a$ corresponds to the intercept of the tangent of the product $r\cdot\psi_0$ at $r=d$ on the $r$-axis in the low-energy limit, while the effective range $r_{\rm eff}$ depends on the range parameter $d$, the depth parameter $V_0$, and the reduced mass $\mu$. For both the shallow and deep bound states, we note that there is a node at $r\simeq a$ (scattering length), which actually determines the suppression of the product $r\cdot\psi_0$ at the long distance.

According to Eq.~\eqref{Eq:CF}, the correlation function depends on two factors, namely, the difference between the free and scattering wave functions  squared, i.e., $\Delta\equiv r^2(|\psi_{l=0}|^2-|j_{l=0}|^2)$, and the source function $S_{12}$, shown as the blue solid line and colored regions in panels (a2-d2) of Fig.~\ref{Fig:NM}. In fact, the above comparison between the free and scattering $r\cdot\psi_0$ can be captured by  the sign of $\Delta$, which is directly related to the properties of the correlation functions in the low-momentum region. In particular,  as the source size $R$ increases, the magnitude of the Gaussian source function decreases rapidly and its tail becomes longer which leads to the reduction of the corresponding correlation function.

The final correlation functions are displayed in panels (a3-d3). The orange solid lines and the green dashed lines denote the results obtained with $R=1$ fm and $R=2$ fm, respectively. From these results, one can conclude that (a) for a repulsive potential the correlation functions are between zero and unity for different $R$; (b) for a weakly attractive potential they are above unity for different $R$; (c) for a moderately attractive potential  the low-momentum correlation function is above unity for small $R$ while below unity for large $R$; and (d) for a strongly attractive potential they are between zero and unity for different $R$. It should be emphasized that the above observations based on the square-well model are consistent with the analysis performed in the Lednicky–Lyuboshitz model~\cite{Lednicky1981YF35.1316, Ohnishi2017PPNP95.279}, but more intuitive.

\section{Summary and outlook}

In this work, we studied the $DK$ correlation function for the first time, which if measured can be used to verify or refute the hadronic molecular picture of $D_{s0}^*(2317)$. We first evaluated the $DK$ coupled-channel interaction in the leading order chiral perturbation theory and calculated the corresponding correlation function. The numerical results showed that the inelastic coupled-channel contribution, which is mainly from the $D^0K^+-D^+K^0$ transition, can be sizable and lead to a cusp-like structure in the $D^0K^+$ correlation function around the $D^+K^0$ threshold. We found that the source size dependence of the $DK$ correlation function is very different from that of moderately strong attractive interactions, which can be utilized to verify the nature of $D_{s0}^*(2317)$ as a deep bound $DK$ state. 

In the next step, based on the square-well model, we studied some general features of correlation functions and explained how one can distinguish  between moderately and strongly attractive interactions via the corresponding correlation functions.

With the large acceptance and the high luminosity upgrade of the ALICE detector~\cite{ALICE2022arXiv2211.02491}, we expect that the $DK$ correlation function can be measured in the near future. The same technique can be utilized to shed light on other hadron-hadron interactions and  the nature of related exotic hadrons.

\begin{acknowledgments} 
We thank Eulogio Oset and the anonymous referee for the many valuable comments which help a lot in refining our presentation. 
This work is partly supported by the National Natural Science Foundation of China under Grant No.11735003, No.11975041, and No. 11961141004, and the fundamental Research Funds for the Central Universities. JXL acknowledges support from the National Natural Science Foundation of China under Grant No.12105006 and China Postdoctoral Science Foundation under Grant No. 2021M690008.
\end{acknowledgments}

\bibliography{DK}

\end{document}